%
\let\includefigures=\iffalse
%
\let\useblackboard=\iftrue
%
%
\newfam\black
\input harvmac.tex
\noblackbox
\includefigures
\message{If you do not have epsf.tex (to include figures),}
\message{change the option at the top of the tex file.}
\def\figin{\epsfcheck\figin}\def\figins{\epsfcheck\figins}
\def\epsfcheck{\ifx\epsfbox\UnDeFiNeD
\message{(NO epsf.tex, FIGURES WILL BE IGNORED)}
\gdef\figin##1{\vskip2in}\gdef\figins##1{\hskip.5in}
\else\message{(FIGURES WILL BE INCLUDED)}%
\gdef\figin##1{##1}\gdef\figins##1{##1}\fi}
\def\DefWarn#1{}
\def\figinsert{\goodbreak\midinsert}
\def\ifig#1#2#3{\DefWarn#1\xdef#1{fig.~\the\figno}
\writedef{#1\leftbracket fig.\noexpand~\the\figno}%
\figinsert\figin{\centerline{#3}}\medskip\centerline{\vbox{\baselineskip12pt
\advance\hsize by -1truein\noindent\footnotefont{\bf Fig.~\the\figno:} #2}}
\bigskip\endinsert\global\advance\figno by1}
\else
\def\ifig#1#2#3{\xdef#1{fig.~\the\figno}
\writedef{#1\leftbracket fig.\noexpand~\the\figno}%
\global\advance\figno by1}
\fi
\useblackboard
\message{If you do not have msbm (blackboard bold) fonts,}
\message{change the option at the top of the tex file.}
\font\blackboard=msbm10
\font\blackboards=msbm7
\font\blackboardss=msbm5
\textfont\black=\blackboard
\scriptfont\black=\blackboards
\scriptscriptfont\black=\blackboardss

\else

\fi
%
\def\yboxit#1#2{\vbox{\hrule height #1 \hbox{\vrule width #1
\vbox{#2}\vrule width #1 }\hrule height #1 }}
\def\fillbox#1{\hbox to #1{\vbox to #1{\vfil}\hfil}}
\def\ybox{{\lower 1.3pt \yboxit{0.4pt}{\fillbox{8pt}}\hskip-0.2pt}}

\def\rightarrowbox#1#2{
  \setbox1=\hbox{\kern#1{${ #2}$}\kern#1}
  \,\vbox{\offinterlineskip\hbox to\wd1{\hfil\copy1\hfil}
    \kern 3pt\hbox to\wd1{\rightarrowfill}}}

\def\comments#1{}

\def\half{{1\over 2}}
\def\Tr{{{\rm Tr}}}

\def\CA{{\cal A}}

\def\CE{{\cal E}}

\def\CN{{\cal N}}

\def\CQ{{\cal Q}}

\def\CV{{\cal V}}

\def\II{\relax{I\kern-.10em I}}

\def\IZ{\relax\ifmmode\mathchoice
{\hbox{\cmss Z\kern-.4em Z}}{\hbox{\cmss Z\kern-.4em Z}}
{\lower.9pt\hbox{\cmsss Z\kern-.4em Z}}
{\lower1.2pt\hbox{\cmsss Z\kern-.4em Z}}\else{\cmss Z\kern-.4em
Z}\fi}
\def\IB{\relax{\rm I\kern-.18em B}}
\def\IC{{\relax\hbox{$\inbar\kern-.3em{\rm C}$}}}
\def\ID{\relax{\rm I\kern-.18em D}}
\def\IE{\relax{\rm I\kern-.18em E}}
\def\IF{\relax{\rm I\kern-.18em F}}
\def\IG{\relax\hbox{$\inbar\kern-.3em{\rm G}$}}
\def\IGa{\relax\hbox{${\rm I}\kern-.18em\Gamma$}}
\def\IH{\relax{\rm I\kern-.18em H}}
\def\II{\relax{\rm I\kern-.18em I}}
\def\IK{\relax{\rm I\kern-.18em K}}
\def\IN{\relax{\rm I\kern-.18em N}}
\def\IP{\relax{\rm I\kern-.18em P}}

%
\def\inbar{\,\vrule height1.5ex width.4pt depth0pt}

\font\cmss=cmss10 \font\cmsss=cmss10 at 7pt
\def\IR{\relax{\rm I\kern-.18em R}}

\def\lp10{l_P^{10}}
\def\lp11{l_P^{11}}
\def\R11{R_{11}}

\def\ra{{\longrightarrow}}
\def\wT{{\widetilde T}}
\def\wG{{\widetilde G}}
\def\wJ{{\widetilde J}}
\def\bz{{\bar z}}
\def\wphi{{\widetilde\phi}}
\def\half{{1\over 2}}
\def\bi{{\bar i}}
\def\bj{{\bar j}}
\def\bk{{\bar k}}
\def\vE{{\check E}}
\def\hom{{\hbox{Hom}}}
\def\pre{{\hbox{Pre-Tr}}}
\def\chom{{\underline{Hom}}}
\newbox\tmpbox\setbox\tmpbox\hbox{\abstractfont IASSNS-HEP/??}
\Title{\vbox{\baselineskip12pt\hbox to\wd\tmpbox{\hss
hep-th/0104200}\hbox{}}}
{\vbox{
\centerline{Enhanced D-Brane Categories from}
\centerline{String Field Theory}}}
\smallskip
\centerline{Duiliu-Emanuel Diaconescu}
\centerline{\it School of Natural Sciences,}
\centerline{\it Institute for Advanced Study,} 
\centerline{\it Princeton, NJ 08540 USA}
\bigskip
\bigskip
\bigskip
\noindent
We construct D-brane categories in ${\bf B}$-type
topological string theory as solutions to string field 
equations of motion. Using the formalism of superconnections,
we show that these solutions form a variant of 
a construction of Bondal and 
Kapranov. This analysis is an elaboration on recent work of  
Lazaroiu. We also comment on the 
relation between string field theory and the derived category 
approach of Douglas, and Aspinwall and Lawrence. 
Non-holomorphic deformations make a somewhat unexpected appearance 
in this construction.
\Date{April 2001}
\def\np{{\it Nucl. Phys.}}

\nref\MRD{M.R. Douglas, 
``D-branes, $\CN=1$ Supersymmetry and Categories'', 
hep-th/0011017}%
\nref\AL{P.S. Aspinwall and A. Lawrence, ``Derived Categories and Zero-Brane 
Stability'', hep-th/0104147.}
\nref\GM{G. Moore, ``Some Comments on Branes, G-flux and K-theory'', 
talk at {\it Strings 2000}, hep-th/0012007.}%
\nref\CLi{C. Lazaroiu, ``On the structure of open-closed topological 
field theory in two dimensions'', hep-th/0010269.}%
\nref\CLii{C. Lazaroiu, ``Generalized complexes and string field 
theory'', 
hep-th/0102122.}%
\nref\CLiii{C. Lazaroiu, ``Unitarity, D-brane dynamics and D-brane
categories'', hep-th/0102183.}%
\nref\CLSP{C. Lazaroiu and S. Popescu, to appear}
\nref\EWi{E. Witten, ``Mirror Symmetry And Topological Field Theory'', 
{\it Mirror Symmetry I}, S.-T. Yau ed., 121, hep-th/9112056.}%
\nref\NW{N.P. Warner, ``$\CN=2$ Supersymmetric Integrable Models 
and Topological Field Theories'', hep-th/9301088.}%
\nref\EWii{E. Witten, ``Chern-Simons Gauge Theory As a String Theory'', 
hep-th/9207094.}%
\nref\EWiii{E. Witten, ``Non-Commutative Geometry And String Field 
Theory'', \np {\bf B268} (1986) 253.}%
\nref\Qi{D. Quillen, ``Superconnections and the Chern Character'', 
Topology {\bf 24} (1985) 89.}%
\nref\CV{C. Vafa, ``Brane/anti-Brane Systems and U(N|M) Supergroup'', 
hep-th/0101218.}%
\nref\BL{J.-M. Bismut and J. Lott, ``Flat Vector Bundles, Direct Images
and Higher Analytic Torsion'', J. Amer. Math. Soc. {\bf 8} (1995) 291.}
\nref\BK{A.I. Bondal and M.M. Kapranov, ``Enhanced Triangulated 
Categories'', 
Math. USSR Sb. {\bf 70} (1991) 93.}%
\nref\Ki{M. Kontsevich, ``Deformation Quantization of 
Poisson Manifolds'', 
q-alg/9709040.}%
\nref\KS{M. Kontsevich and Y. Soibelman, ``Homological 
mirror symmetry and torus fibrations'', math.SG/0010041.}%
\nref\RPTa{R.P. Thomas, Ph. D. thesis.}%
\nref\RPTb{R.P. Thomas, ``A Holomorphic Casson Invariant for Calabi-Yau 
Threefolds And Bundles on $K3$ Fibrations'', alg-geom/9806111.}%
\nref\OOY{H. Ooguri, Y. Oz and Z. Yin, 
``D-Branes on Calabi-Yau Spaces and Their Mirrors'', 
Nucl. Phys. {\bf B477} (1996) 407, hep-th/9606112.}%
\nref\KHi{K. Hori, ``D-Branes, T-duality, and Index Theory'', 
ADTMP {\bf 3} (1999) 281, hep-th/9902102.}
\nref\EWiv{E. Witten, ``Overview of K-Theory Applied to Strings'', 
hep-th/0007175.}
\nref\SZ{A. Sen, B. Zwiebach, 
``Tachyon condensation in string field theory'', 
JHEP {\bf 03} (2000) 002, hep-th/9912249.}%
\nref\RSZ{L. Rastelli, A. Sen, and B. Zwiebach, 
`` String Field Theory Around 
The Tachyon Vacuum'', hep-th/0012251; 
``Classical Solution in String Field 
Theory Around The Tachyon 
Vacuum'', hep-th/0102112.}%
\nref\MZ{J.A. Minahan, B. Zwiebach, 
``Effective Tachyon Dynamics in Superstring
Theory'', hep-th/0009246.}%
\nref\MSZ{N. Moeller, A. Sen and B. Zwiebach, 
``D-branes as Tachyon Lumps in 
String Field Theory'', JHEP {\bf 08} (2000) 039, hep-th/0005036.}%
\nref\NSZ{N. Berkowitz, A. Sen, and B. Zwiebach, 
``Tachyon Condensation in 
Superstring Field Theory'', Nucl. Phys. {\bf B587} (2000) 147, 
hep-th/0002211.}%
\nref\B{N. Berkowitz, ``The Tachyon Potential in Open 
Neveu-Schwarz String Field Theory'', 
JHEP {\bf 04} (2000) 022, hep-th/0001084.}%
\nref\KJMT{R. de Mello Koch, A. Jevicki, M. Mihailescu, and R. Tatar, 
`` Lumps and P-branes in Open String Field Theory'', 
Phys. Lett. {\bf B482} (2000) 249, hep-th/0003031.}%
\nref\WT{W. Taylor, 
``D-Brane Effective Action From String Field Theory'', 
Nucl. Phys. {\bf B585} (2000) 171, hep-th/0001201.}%
\nref\GS{A.A. Gerasimov, S.L. Shatashivili, ``On Exact Tachyon Potential 
in Open String Field Theory'', JHEP {\bf 10} (2000) 034, 
hep-th/0009103.}%
\nref\KMM{D. Kutasov, M. Marino, and G. Moore, ``Some Exact Results on 
Tachyon Condensation in String Field Theory'', JHEP {\bf 10} (2000) 045,
hep-th/0009148; ``Remarks on Tachyon Condensation in Superstring 
Field Theory'', hep-th/0010108.}%
\nref\DMR{K. Dasgupta, S. Mukhi, G. Rajesh, ``Noncommutative Tachyons'', 
JHEP {\bf 06} (2000) 022, hep-th/0005006.}%
\nref\HKLM{J. A. Harvey, P. Kraus, F. Larsen, and E. J. Martinec, 
``D-branes and Strings as Non-commutative Solitons'',
JHEP {\bf 07} (2000) 042, hep-th/0005031.}%
\nref\TTU{T. Takayanagi, S. Terashima, T. Uesugi,
``Brane-Antibrane Action from Boundary String Field Theory'',
JHEP {\bf 03} (2001) 019, hep-th/0012210.}%
\nref\KL{P. Kraus, F. Larsen, 
``Boundary String Field Theory of the DDbar System'', hep-th/0012198.}%
\nref\dA{S.P. de Alwis, ``Boundary String Field Theory, the 
Boundary State Formalism and D-Brane Tension'', hep-th/0101200.}%
\nref\RS{A. Recknagel, V. Schomerus, ``D-branes in Gepner models'', 
Nucl. Phys. {\bf B531} (1998) 185, hep-th/9712186.}%
\nref\GS{M. Gutperle, Y. Satoh, ``D-branes in Gepner models and 
supersymmetry'', Nucl. Phys. {\bf B543} (1999) 73
hep-th/9808080.}
\nref\BDLR{I. Brunner, M. R. Douglas, A. Lawrence, and 
C. Romelsberger, ``D-branes on the Quintic'', JHEP {\bf 08} (2000) 015,
hep-th/9906200.}%
\nref\DFRi{M. R. Douglas, B. Fiol, C. Romelsberger, 
``Stability and BPS branes'', hep-th/0002037.}%
\nref\DFRii{M. R. Douglas, B. Fiol, C. Romelsberger,
``The spectrum of BPS branes on a noncompact Calabi-Yau'', 
hep-th/0003263.}%
\nref\DD{D.-E. Diaconescu, and M.R. Douglas, 
``D-branes on Stringy Calabi-Yau Manifolds'', hep-th/0006224.}%
\nref\M{P. Mayr, ``Phases of Supersymmetric D-branes on Kaehler 
Manifolds and the McKay correspondence'', JHEP {\bf 01} (2001) 018,
hep-th/0010223.}%
\nref\RPH{R.P. Horja, ``Hypergeometric functions and mirror 
symmetry in toric varieties'', math.AG/9912109; 
``Derived Category Automorphisms from Mirror Symmetry'', 
math.AG/0103231.}%
\nref\LMW{W. Lerche, P. Mayr, J. Walcher, 
``A new kind of McKay correspondence from non-Abelian gauge theories'',
hep-th/0103114.}%
\nref\GJ{S. Govindarajan, T. Jayaraman, ``On the Landau-Ginzburg 
description of Boundary CFTs and special Lagrangian submanifolds'', 
JHEP {\bf 07} (2000) 016, hep-th/0003242;
``D-branes, Exceptional Sheaves and Quivers on Calabi-Yau manifolds: 
From Mukai to McKay'', hep-th/0010196; ``Boundary Fermions, 
Coherent Sheaves and D-branes on Calabi-Yau manifold'', hep-th/0104126.}%
\nref\GJS{S. Govindarajan, T. Jayaraman, and T. Sarkar, 
``On D-branes from Gauged Linear Sigma Models'', Nucl. Phys. {\bf B593} 
(2001) 155, hep-th/0007075.}%
\nref\DG{D.-E. Diaconescu, J. Gomis, 
``Fractional Branes and Boundary States in Orbifold Theories'', 
hep-th/9906242.}%
\nref\DR{D.-E. Diaconescu, C. Romelsberger, 
``D-Branes and Bundles on Elliptic Fibrations'', 
Nucl. Phys. {\bf B574} (2000) 245, hep-th/9910172.}%
\nref\KLLW{P. Kaste, W. Lerche, C.A. Lutken, J. Walcher, 
``D-Branes on K3-Fibrations'', Nucl. Phys. {\bf B582} (2000) 
203, hep-th/9912147.}%
\nref\HIV{K. Hori, A. Iqbal, C. Vafa, ``D-Branes And Mirror Symmetry'', 
hep-th/0005247.}%
\nref\Hi{K. Hori, ``Linear Models of Supersymmetric D-Branes'', 
hep-th/0012179.}%
\nref\ES{E. Scheidegger, ``D-branes on some 
one- and two-parameter Calabi-Yau hypersurfaces'', 
JHEP {\bf 04} (2000) 003, 
hep-th/9912188.}
\nref\GL{B. Greene, C. Lazaroiu, ``Collapsing D-Branes in 
Calabi-Yau Moduli Space: I'', hep-th/0001025.}%
\nref\SY{K. Sugiyama, S. Yamaguchi, 
``D-branes on a Noncompact Singular Calabi-Yau Manifold'', 
JHEP {\bf 02} (2001) 015, hep-th/0011091.}
\nref\ACY{B. Andreas, G. Curio, D. Hernandez Ruiperez, S.-T. Yau,
``Fourier-Mukai Transform and Mirror Symmetry for D-Branes on 
Elliptic Calabi-Yau'', math.AG/0012196; 
``Fibrewise T-Duality for D-Branes on Elliptic Calabi-Yau'', 
JHEP {\bf 03} (2001) 020, hep-th/0101129.}%
\nref\Ai{P.S. Aspinwall, ``Some Navigation Rules for 
D-Brane Monodromy'', 
hep-th/0102198.}%
\nref\KMc{S. Kachru, J. McGreevy, 
``Supersymmetric Three-cycles and (Super)symmetry Breaking'', 
Phys. Rev. {\bf D61} (2000) 026001, hep-th/9908135.}%
\nref\KKLM{S. Katz, S. Kachru, A. Lawrence, and J. McGreevy, 
`` Open string instantons and superpotentials'', 
Phys. Rev. {\bf D62} (2000) 026001, hep-th/9912151; 
Mirror symmetry for open strings'', Phys. Rev. {\bf D62} (2000) 126005, 
hep-th/0006047.}%
\nref\BF{B. Fiol, ``The BPS Spectrum of N=2 SU(N) SYM and Parton 
Branes'', 
hep-th/0012079.}%
\nref\BRG{B.R. Greene, ``Aspects of Collapsing Cycles'', 
hep-th/0011059.}%
\nref\KO{A. Kapustin and D. Orlov, 
``Vertex Algebras, Mirror Symmetry, And D-Branes: The Case Of 
Complex Tori'', 
hep-th/0010293.}%
\nref\CLiv{C. Lazaroiu, 
``Instanton amplitudes in open-closed topological string theory'', 
hep-th/0011257.}%
\nref\MOY{K. Mohri, Y. Onjo, S.-K. Yang, ``Closed Sub-Monodromy 
Problems, 
Local Mirror Symmetry and Branes on Orbifolds'', hep-th/0009072.}%
\nref\BV{I. Brunner, V. Schomerus, 
``On Superpotentials for D-Branes in Gepner Models'', 
hep-th/0008194.}%
\nref\CLv{C. Lazaroiu, 
``Collapsing D-branes in one-parameter models and small/large 
radius duality'',  hep-th/0002004.}%
\nref\MR{M. Raugas, 
``D-Branes and Vanishing Cycles in Higher Dimensions'', 
hep-th/0102133.}%
\nref\BD{I. Brunner, J. Distler, 
``Torsion D-Branes in Nongeometrical Phases'', hep-th/0102018.}%
\newsec{Introduction} 

D-brane dynamics has been the object of much recent attention. 
There are roughly two main current directions of research in this 
area. Many of the recent papers are concerned with a fundamental 
microscopic description of D-branes by means of tachyon dynamics in 
open string field theory \refs{\SZ-\dA}. 
On the other hand, considerable effort has 
been made in order to improve our understanding of low energy 
effective dynamics of D-branes in situations with $N=1$ supersymmetry
\refs{\RS-\BD}.
In particular, Douglas \MRD\ has proposed a beautiful formal structure 
underlying D-branes in topologically twisted $N=(2,2)$ superconformal 
field theories. According to his work, and also to the detailed 
analysis of Aspinwall and Lawrence \AL,\ we have to revise 
our traditional understanding of supersymmetric (even) branes on 
Calabi-Yau manifolds. Very briefly, they showed that D-branes are 
properly thought of as objects in a special category associated to 
a Calabi-Yau space $X$ -- the bounded derived category of coherent 
sheaves $D^b(X)$. In down to earth terms, this amounts to including 
differential complexes of coherent sheaves among physical D-branes. 

In remarkable parallel work, Lazaroiu \refs{\CLi,\CLii,\CLiii} 
has developed a very 
general approach to D-branes in string field theory. Using unitarity 
constraints and very general string field considerations, he has shown 
that D-branes naturally form certain enlarged 
categories equipped with special structures (such as a differential 
graded structure.) An axiomatic approach to topological open-closed 
string theories has been discussed in 
\refs{\GM, \CLi}.

The purpose of the present work is to study in more detail
the D-brane category in topological string field theory, and 
to investigate the relation with the derived category of 
\refs{\MRD, \AL}.  
Some elements along these lines 
have been sketched in \refs{\CLii, \CLiii}. We take a pragmatic
approach, by constructing an extension of Witten's holomorphic 
Chern-Simons theory \EWii.\ The crucial element in this approach is the 
$\IZ$ grading of topological boundary states introduced in \MRD. 
This allows us to effectively identify the lowest mode expansion 
of the string field as a (graded) superconnection \Qi. It turns out 
that the cubic topological string field theory reduces to a Chern-Simons 
theory for superconnections, using arguments similar to \EWii.
Note that superconnections have appeared in a similar 
context in \CV.\ 
The solutions to the string field equations of motion are closely 
related to the twisted complexes defined by Bondal and Kapranov 
\BK. Very briefly, these are 
collections $\{E_n\}$ of holomorphic bundles with 
``maps'' $q_{mn}$ between various (non-consecutive) $E_m, E_n$ 
satisfying a Maurer-Cartan equation. 
The precise definition of such objects will be given in sections three 
and four. The associated categorical structure has been constructed 
by Bondal and Kapranov in \BK. According to the proposal of 
\refs{\CLii, \CLiii}, one should construct a more 
general category satisfying a certain completion condition.
\foot{I thank C. Lazaroiu for pointing this out. See also 
\refs{\CLSP}.} 
We can avoid performing such a construction 
by restricting to the particular class of solutions described above. 

We conclude that the class of topological open string theories 
considered in this paper form a variant of a Bondal-Kapranov category. 
A generalized D-brane, i.e. an object in this 
category is a twisted complex, as sketched above. Perhaps 
one of the most striking aspects of this analysis is that, although 
we start with an open string background defined by holomorphic vector 
bundles $E_n$, we soon find general solutions of string field 
theory based on non-holomorphic deformations of the $E_n$. 
Nevertheless these determine consistent topological open string theories,
with a good fermionic symmetry. Such solutions could be 
described as holomorphic superconnections.

A legitimate question is if these solutions define new topological 
branes 
or they are just artifacts of the string field approach. 
More specifically, 
one would like to know what is the relation between these 
D-brane categories 
and the derived categories found in \refs{\MRD, \AL}. 
This is an interesting 
question, but we can provide only a partial answer in section four. 
We show 
that the derived category is equivalent to a full subcategory of the 
D-brane category, by a careful comparison with \AL. 
However, we are unable to settle the question if these two 
categories are equivalent in spite of the apparent differences.
This would prove that the string field approach brings nothing new. 
We expect that solving this puzzle would involve an alternative 
formulation of the string field category, perhaps in pure algebraic 
terms, if such a formulation exists. This is likely to be related 
to the approach of Kontsevich \Ki\ in the context of homological 
mirror symmetry (see also \KS.)

\newsec{The Topological {\bf B} Model} 

This is standard material, so we will review only what is needed. 
Recall that the standard $N=(2,2)$ superconformal algebra is 
generated by a set 
$T(z), G^{\pm}(z), J(z)$, of holomorphic currents, and a similar set 
${\wT}(\bz), \wG^\pm(\bz), 
\wJ(\bz)$ of anti-holomorphic currents. 
It is common practice to introduce a bosonic representation of the 
$U(1)$ current 
\eqn\bosonic{
J(z) = i\sqrt{{\hat c}}\del \phi(z),\qquad
\wJ(\bz) = i\sqrt{{\hat c}}{\bar \del}\wphi(\bz).}
If the theory is formulated on the half-plane, 
one can impose either ${\bf A}$-type of ${\bf B}$-type 
boundary conditions preserving $\CN=2$ superconformal symmetry 
\OOY.\
In this paper, we will be exclusively concerned with ${\bf B}$ boundary 
conditions 
\eqn\bc{
G(z)^\pm = \wG^\pm(\bz),\qquad J(z) = \wJ(\bz).}
Note that the second equation is equivalent to Neumann 
boundary conditions for the compact boson $\phi(z,\bz)$.

Now let us discuss topological twists \EWi,\ ffollowing closely 
\NW. The main point is to alter the energy momentum tensor 
\eqn\toptwist{\eqalign{
& T(z) \ra T(z)_{top} = T(z) \pm \half \del J(z)\cr
& \wT(\bz) \ra \wT(\bz)_{top} = \wT(\bz) \pm 
\half {\bar \del}\wJ(\bz).\cr}}
Although it looks as if we have many choices, only the relative 
sign between the holomorphic and the anti-holomorphic part is 
relevant. This yields two types of topological string models 
dubbed again type ${\bf A}$, when the signs are opposite and 
type ${\bf B}$, when the signs are the same. To fix conventions, 
we will always take the sign of the holomorphic twist to be plus. 
The main effect is a shift of the conformal weight of all operators 
in the theory 
\eqn\confshift{
h\ra h_{top} = h -\half q,}
where $q$ is the $U(1)$ charge. The supercharge $G^+_{-\half}$ becomes 
a nilpotent BRST charge $Q$ in the twisted theory, $Q^2=0$. 
Accordingly, the $U(1)$ charge becomes ghost charge, and the $U(1)$ 
current $J(z)$ is simply the ghost number operator. 

Next we consider open-closed topologically twisted models on the 
half-plane. By inspecting 
\bc,\ \toptwist,\ it is clear that a ${\bf B}$ twist is compatible only 
with ${\bf B}$ boundary conditions, and this will be the case considered
in this paper. 
This theory has been analyzed in the context of nonlinear sigma 
models with Calabi-Yau target space in \refs{\EWii}. Let us denote 
by $X$ the Calabi-Yau manifold.  The main result of \EWii\ 
is that the open string sigma 
model can be consistently coupled to a holomorphic bundle $E$ on $X$, 
and the cubic string field theory action reduces in this case to a 
holomorphic Chern-Simons gauge theory. 

In order to facilitate the presentation, it may be helpful to 
recall some details of the analysis of \EWii. Let $\Phi:\Sigma\ra 
X$ denote the map from the string world-sheet to the target space $X$. 
Recall \refs{\EWi, \EWii} 
that the fermi fields of the topological {\bf B} model are $\eta^{\bi}, 
\theta^{\bi}$ sections of $\Phi^\ast(T^{0,1}(X))$ and $\rho^i$ which is a
section of $T^\ast(\Sigma)\otimes \Phi^\ast(T^{1,0}(X))$. We also define 
$\theta_j = g_{j\bi}\theta^{\bi}$. 
The Lagrangian of the {\bf B} model is 
\eqn\Blagr{\eqalign{
L = t\int_\Sigma d^2z & \bigg(g_{i\bj}\del_z\phi^i\del_\bz \phi^{\bj} 
+ i \eta^{\bi}(D_z\rho_{\bz}^i+D_{\bz}\rho^i_z)g_{i\bi}\cr
&+i\theta_i(D_\bz\rho_z^i-D_z\rho_\bz^i) + R_{i\bi j\bj} 
\rho^i_z\rho^j_\bz\eta^\bi\theta_kg^{k\bj}\bigg),\cr}}
which is invariant under the following fermionic symmetry
\eqn\infBRST{\eqalign{
& \delta \phi^i = 0\cr
& \delta \phi^{\bi} = i \epsilon \eta^{\bi}\cr
&\delta \eta^{\bi} = \delta \theta_i = 0\cr
& \delta \rho^i = -\epsilon d\phi^i.\cr}}
This model can be coupled to a background gauge field $A$ on $X$ via the 
boundary coupling
\eqn\bdterm{
L_{bdry} = \int_{\del \Sigma} \Phi^*(A)-\eta^{\bi}F_{\bi j} \rho^j.}
It has been shown in \EWii\ that this coupling preserves the fermionic 
symmetry if and only if 
\eqn\presfermi{
F_{\bi \bj} (A) = 0.}
This means that the operator 
\eqn\dolbop{
{\bar \del}_A={\bar \del} + A^{0,1}}
defines an integrable holomorphic structure on the gauge bundle $E$. 
In the following we will fix such a background holomorphic bundle 
$E$ , and we will adopt the notation ${\bar \del}_E$ for the covariant 
Dolbeault operator.

Using standard string field theory arguments \refs{\EWii}, 
the physical states can be found by computing the cohomology of $Q$ 
on the kernel of the Hamiltonian $L_0$ derived from \Blagr.\
This has been done in \EWii,\ for the case at hand
with the result that in the $t\ra \infty$ limit the eigenfunctions 
localize on the subspace of constant maps $\Phi: I \ra X$. 
Making use of the canonical commutation relations 
for fermions, we can write the 
string field as a wave functional depending on the zero modes
of $\phi^I$ and $\eta^\bi$
\eqn\sfA{
\Psi(\phi^I, \eta^\bi) = a^0(\phi^I) + \eta^\bi a^1_\bi(\phi^I) 
+\eta^\bi\eta^\bj a^2_{\bi\bj}(\phi^I) + \eta^\bi\eta^\bj\eta^\bk 
a^3_{\bi\bj\bk}(\phi^I) + \ldots}
The components of the string field can be identified with $(0,q)$ 
forms on $X$ where the degree $q$ is the ghost number. 
The BRST operator $Q$ in the background $A$ 
can be shown to correspond to the Dolbeault operator 
$\bar\del_E$, hence the physical states correspond 
to cohomology classes in $H^{0,q}(E^\ast\otimes E)$. 

Since the degree $q$ coresponds to the ghost charge, 
the only ghost number 
one term in \sfA\ is the linear piece 
$a^1\in \Omega^{0,1}(X, E^\ast\otimes E)$.
This is to be interpreted as a deformation of the operator 
$\bar\del_E$. 
Then, one can show \EWii\ that the cubic string field action 
reduces in this 
case to holomorphic Chern-Simons theory 
\eqn\holCS{
S= \half \int_X \Omega\wedge \hbox{Tr} 
\left(a^1\wedge {\bar \del}_E a^1 + {2\over 3} 
a^1\wedge a^1\wedge a^1\right),}
where 
$\Omega$ is a nonvanishing holomorphic three-form on $X$. 
The equation of motion derived from this action 
is 
\eqn\eomA{
{\bar \del}_E a^1 + a^1\wedge a^1 = 0.}
which means that the deformed operator ${\bar\del}_A + a^1$ is again 
integrable, therefore it defines a new holomorphic structure on $E$. 
 Moreover, the gauge transformations 
of string field 
theory reduce in the $t\ra \infty$ to 
ordinary complex gauge transformations 
\eqn\gauge{
a^1 \ra a^1 + {\bar \del}_E \epsilon.}
Two deformations related by a gauge 
transformation give rise to isomorhic 
complex structures. It follows that the target space action of 
{\bf B} topological field theory is intimately related to deformation 
of holomorphic bundles \refs{\RPTa,\RPTb}.

To conclude this section, let us reformulate the above considerations 
in categorical language \refs{\CLii,\CLiii}.
The differential graded category of
off-shell open string states can be defined as follows 
\refs{\CLii,\CLiii}.
The objects consist of holomorphic vector bundles $E$ over $X$. 
Given two objects $E, F$, we define 
\eqn\morphisms{
\hbox{Hom}_\CE(E,F) = \oplus_{q=0}^3\Omega^{0,q}(E^\ast\otimes F).}
Note that $\hbox{Hom}_\CE(E,F)$ 
has a natural structure of graded abelian group, 
with the grading given by the ghost charge $q$. 
We also define 
a differential 
\eqn\differential{
d_\CE : \hbox{Hom}^q_\CE(E,F)\ra \hbox{Hom}^{q+1}_\CE(E,F), \qquad 
d_\CE = {\bar \del}_{E^\ast\otimes F},}
making $\hbox{Hom}_\CE(E,F)$ a differential complex.
The associated 
cohomology category $H(\CE)$ is defined as having the same objects,
and morphisms given by 
\eqn\cohcateg{
\hbox{Hom}_{H(\CE)}(E,F) = H(\hbox{Hom}^{\bullet}_\CE(E,F)).}
One can also truncate consistently to cohomology in zero degree 
in \cohcateg,\ obtaining another cohomology category $H^0(\CE)$. 
The physical intepretation should be clear: the objects represent 
topological D-branes, while the morphisms represent the off-shell 
open string states in the presence of two D-branes $E$ and $F$. 
The differential is the BRST operator which defines physical states.
Passing to the cohomology category is equivalent to keeping 
only physical open string states.  
We will see later that if we introduce a grading of boundary states, 
we will find a much bigger D-brane category. 

\newsec{Grading and a Generalization of Holomorphic Chern-Simons Theory}

In this section we present a generalization of the previous analysis 
which takes into account the $\IZ$ grading of boundary states discovered 
in \refs{\MRD}. This is one of the main ingredients of \MRD\ 
in establishing the relation between boundary states in topological 
models and derived categories. 
In order to explain the main idea, recall that the theory contains a 
$U(1)$ current $J= i\sqrt{{\hat c}}\del \phi$ which becomes ghost 
number operator 
in the topological model. For ${\bf B}$ models, the compact boson
$\phi$ is subject to Newmann boundary conditions. This means that the 
open string states stretching between two D-branes $E,F$ will carry a
quantum number representing KK momentum around the circle. In the 
topological theory, this quantum number is a boundary ghost charge. 
For {\bf A} models, the same quantum number has been 
described as winding number \MRD.\ 
Since we are working in off-shell string field theory, the effect of 
this quantum number is to induce a grading on the space of boundary 
states. In other words, we have to distinguish between a D-brane 
$E$ and a D-brane defined by an isomorphic holomorphic bundle 
if there is an open string with boundary ghost charge $p$ stretching 
between them. 
More concretely, this means that a D-brane must be specified 
by a holomorphic 
Chan-Paton bundle on $X$ together with an integer $n\in\IZ$. 
The open string states 
stretching between the D-branes $E_n$ and $E_{n+p}$ will carry $p$ units 
of ghost charge. 

Given these considerations we can now proceed with the analysis of 
string field theory in the background of a graded collection of 
D-branes $\{E_n\}$. The $E_n$ are holomorphic vector bundles over 
$X$. Note that this is not the most general configuration possible 
since one assumes that the bundle $E_n$ has grade $n$. 

First, 
we write down the most general expansion of the 
string field 
\eqn\sfB{
\Psi(\phi^I, \eta^\bi)=\sum_{m,n}a^0_{mn} +
\eta^{\bi}(a^1_\bi)_{mn} + \eta^\bi\eta^\bj(a^2_{\bi\bj})_{mn} 
+ \eta^\bi\eta^\bj\eta^\bk(a^3_{\bi\bj\bk})_{mn},}
where $c_{mn}$ are maps from $E_m$ to $E_n$ i.e. sections 
of $E_m^\ast\otimes E_n$. Similarly, the higher order terms 
can be regarded as sections of $\Omega^{0,q}(X)\otimes 
(E_m^\ast\otimes E_n)$. 
Keeping in mind the relation between grading and ghost number, 
it follows 
that a section of $\Omega^{0,q}(X)\otimes(E^\ast_m\otimes E_n)$, 
has ghost number $q+(n-m)$. Therefore the ghost number one piece of 
$\Psi(\phi^I,\eta^\bi)$ is 
\eqn\sfC{
\Psi^{(1)}(\phi^I,\eta^\bi)=\sum_{n}a^0_{n,n+1}+
\eta^{\bi}(a^1_\bi)_{n,n} + \eta^\bi\eta^\bj(a^2_{\bi\bj})_{n+1,n} 
+ \eta^\bi\eta^\bj\eta^\bk(a^3_{\bi\bj\bk})_{n+2,n}.}
The next step is to compute the cubic string field action in terms 
of the components of \sfC.  

At this stage it may be helpful to discuss some of the mathematical 
structure 
underlying open string field theory
\refs{\EWiii}. 
The string fields form 
an associative noncommutative graded algebra $\CA$, 
the grading being defined 
by the ghost number. This algebra is endowed with a derivation $Q$, 
which is the BRST operator, and with a trace map $\int: \CA \ra \IC$, 
satisfying the following rules 
\eqn\rules{\eqalign{
&Q(a\ast b) = (Qa) \ast b +(-1)^{\deg(a)}a\ast (Qb)\cr
&\int a\ast b = (-1)^{\deg(a)\deg(b)}\int b \ast a\cr
&\int Qa = 0.\cr}}
The structure in \rules\ defines a differential graded 
algebra.\foot{The axioms \rules\ do not 
suffice to describe a theory with D-branes. As discussed in \CLii,\
what is missing is a category structure. It is important to check 
that the string field product and metrics decompose in a manner
consistent with this structure.}

Following the strategy of \EWii,\ we first give a more concrete 
description 
of this algebra for topological open strings in the large $t$ limit. 
In this case, the string field admits an expansion in terms of 
lowest lying modes which can be identified with elements of 
$\Omega^{0,q}(E_m^\ast\otimes E_n)$. Therefore the underlying 
space of the algebra 
would naively be identified with 
\eqn\algebraA{
\oplus_{q=0}^3 \oplus_{(m,n) \in \IZ^2}  
\Omega^{0,q}(E_m^\ast\otimes E_n),}
with the grading defined by the ghost number $q+(n-m)$. 
In fact, it turns out that working with $\IZ$-graded bundles is not 
quite enough in order to reproduce the structure of the 
string field algebra $\CA$. 
In order to reproduce the relations \rules\ the naive proposal \algebraA\
must be refined by working with $\IZ$-graded 
{\it super} vector bundles. In other words, each graded vector 
bundle $\{E_n\}$ can be viewed as a graded super vector bundle 
$\{\vE_n\}$, where 
\eqn\super{
\vE_n=\left\{\matrix{(E_n, 0)\qquad & \hbox{for}\ n\ \hbox{even}\cr
& \cr
(0, E_n)\qquad & \hbox{for}\ n\ \hbox{odd}.\cr}\right.}
This is a standard construction \Ki.
Given such an object, we can obtain either a $\IZ$-graded bundle by forgetting 
the $\IZ/2$ grading or a super vector bundle by forgetting the $\IZ$-grading. 
In the last case, the resulting $\IZ/2$-graded bundle $\vE=(E^+, E^-)$
has components 
\eqn\superB{
E^+ = \oplus_{k}E_{2k},\qquad E^-=\oplus_{k}E_{2k+1}.}
Now, note that we have a superalgebra 
$\Omega(X) = \oplus_{q=0}^3 \Omega^{0,q}(X)$ (with standard 
multiplication of forms), and another superalgebra
$End(\vE)= \oplus_{(m,n)\in \IZ^2}\Omega^{0,0}(E_m^\ast\otimes E_n)$. 
In the second case, there is an extra $\IZ$ grading defined by $(n-m)$; 
if we ignore this grading, the superalgebra structure is the standard 
one \Qi. 
Now we can construct the $\IZ$-graded superalgebra 
\eqn\superC{
\CA = \Omega(X)\otimes_{\Omega^{0,0}(X)} \hbox{End}(\vE).}
In order to keep track of various gradings, we introduce the 
notation $\CA^q_{(m,n)}= \Omega^{0,q}(E_m^\ast\otimes E_n)$, 
so that we have 
\eqn\decomp{
\CA = \oplus_{m,n,q} \CA^q_{(m,n)}.}
The degree of an element $f\in \CA^q_{m,n}$ is 
$\hbox{deg}(f) = q+(n-m)$.
Again, if we ignore the $\IZ$-grading, this is the standard 
tensor product of superalgebras \Qi. If $\omega,\eta\in \Omega(X)$, 
and $f,g\in \hbox{End}(\vE)$, we have 
\eqn\superD{
(\omega\otimes f) (\eta\otimes g) = 
(-1)^{\hbox{deg}(f)\hbox{deg}(\eta)}
(\omega\wedge\eta) (fg).}
A similar construction can be found in a different context in \BL.
 
We claim that this is the correct construction of the 
topological open string algebra. 
Let us describe the remaining elements. 
The trace map $\int:\CA\ra \IC$ is given by 
\eqn\tracemap{
\int f = \int \Omega\wedge \Tr_s(f),}
where $\Tr_s:\CA\ra\Omega(X)$ 
denotes the supertrace of \Qi. It is a standard fact that 
\eqn\superE{
\Tr_s(fg) = (-1)^{\deg(f)\deg(g)}\Tr_s(g f)}
for any two elements $f,g\in \CA$. 
The BRST operator is a superconnection $D: \CA\ra \CA$ satisfying the 
Leibniz rule \Qi\
\eqn\leibniz{
D(\omega f) = {\bar\del}\omega + (-1)^{\deg(\omega)} \omega Df}
for all $\omega\in \Omega(X)$, $f\in \CA$. In the particular case 
under study, $D$ has the special form 
\eqn\superF{
D=\oplus_{n\in \IZ} ({\bar\del}_{E_n}).} 
Given, \superE,\ and \leibniz,\ one can check that the relations 
\rules\ are satisfied.

We are now ready to write down the cubic action of topological 
open string field theory. Recall that we have to consider only the 
ghost number one piece which is reproduced below for convenience
\eqn\ghone{
\Psi^{(1)} = \sum_{n} a^0_{n,n+1} + a^1_{nn} + a^2_{n+1, n} + a^3_{n+2, n},}
where $a^q_{m,n}\in \CA^q_{m,n}$. Using the graded superalgebra 
structure 
discussed so far, the cubic action can be written in compact form
 \eqn\cubicC{
S = \half \int_X \Omega\wedge\Tr_s\left(\Psi^{(1)}D\Psi^{(1)} + 
{2\over 3} \Psi^{(1)}\Psi^{(1)}\Psi^{(1)}\right).}
This is the super extension of the holomorphic Chern-Simons action 
mentioned in the introduction. Substituting \ghone\ into \cubicC\ 
we obtain the following expression
\eqn\cubicB{\eqalign{
S = \half \int_X &\Omega\wedge\bigg[
\Tr_s\left(a^1(Da^1) + a^0(Da^2) + a^2(Da^0)\right)+\cr
& {2\over 3}\Tr_s\left(a^1a^1a^1+ 
a^0a^0a^3+a^0a^3a^0+ a^3a^0a^0\right)+\cr
&{2\over 3}\Tr_s\left(a^0a^1a^2+ (\hbox{ all permutations})
\right)\bigg],
\cr}}
where $a^q$ is a sum over all $n$. For example 
\eqn\compnot{
a^0=\sum_{n} a^0_{n, n+1}}
and so on. 
Reasoning by analogy with the ungraded case, we can view $\Psi^{(1)}$ as 
a deformation of the superconnection $D$ defined in \superF. Since the 
bundles $E_n$ that we started with are holomorphic, D satisfies the 
integrability condition $D^2=0$. We will refer to this condition 
as flatness. The equations of motion derived from \cubicC\ read 
\eqn\eqmotionA{
D\Psi^{(1)} + \Psi^{(1)}\Psi^{(1)}=0,}
or, in components, 
\eqn\eqmotionB{\eqalign{
& a^0_{n+1, n+2} a^0_{n, n+1} = 0 \cr
& D a^0_{n, n+1} + a^0_{n, n+1} a^1_{n,n} + a^1_{n+1, n+1} 
a^0_{n, n+1}= 0\cr
& Da^1_{n, n} + a^1_{n,n} a^1_{n,n} + a^2_{n+1, n} a^0_{n, n+1} 
+ a^0_{n-1,n}a^2_{n, n-1} = 0\cr
& Da^2_{n+1, n} + a^1_{n,n}a^2_{n+1,n}+ a^2_{n+1,n} a^1_{n+1, n+1}+
 a^3_{n+2,n}a^0_{n+1, n+2} + a^0_{n-1,n} 
a^3_{n+1, n-1} = 0.\cr}}
An important point is that these equations are equivalent to flatness 
of the deformed superconnection $D+\Psi^{(1)}$. 
Therefore on-shell string field configurations are 
determined by flat superconnections of a general form (as opposed 
to $D$, which is a diagonal superconnection.) Formulated differently, 
the above argument shows that, given a topological open string theory 
defined by the collection of D-branes $\{E_n\}$, we can deform 
by arbitrary operators $a^0_{n, n+1}, a^1_{n,n},\ldots$
One obtains consistent 
topological open string theories as long as the flatness conditions 
\eqmotionA,\ \eqmotionB\ are satisfied. 

It is interesting to note that the equations \eqmotionB\ do not enforce 
holomorphic deformations of the bundles $E_n$. 
Namely, consider the 
second equation in \eqmotionB\
\eqn\flatnessA{
Da^1_{n, n} + a^1_{n,n} a^1_{n,n} + a^2_{n+1, n} a^0_{n, n+1} 
+ a^0_{n-1,n}a^2_{n, n-1} = 0.}
$a^1_{n,n}$ is a deformation of the covariant Dolbeault operator 
${\bar \del}_{E_n}$. As noted before, this deformation defines a 
new holomorphic structure if and only if 
\eqn\holdef{
{\bar \del}_{E_n}a^1_{n, n+1} + a^1_{nn} a^1_{nn} = 0.}
Therefore the equation \flatnessA\ allows nonholomorphic deformations 
of the $E_n$ at the price of exciting the higher $q$-form fields 
$a^q$, $q\geq 2$. We do not know at this stage if these are genuine 
new deformations of the topological {\bf B} model. For example, we can 
turn values of the higher fields $a^q$ such that holomorphy is
preserved. These would correspond to the on-shell deformations 
of \refs{\MRD, \AL}, where it has been shown that they do not 
give anything new beyond the derived category. The problematic 
deformations are the non-holomorphic ones.
This can be hopefully 
settled by searching for examples in concrete models, and we leave this 
for future work. 

In the following we try to elucidate the categorical structure of the 
D-branes found above and comment on the relation with 
derived categories. 

\newsec{Twisted Complexes and Enhanced Triangulated Categories} 

In this section we will show that the solutions to string field theory 
found above form a category $\CQ$ closely related to 
the enhanced triangulated categories defined 
by Bondal and Kapranov \BK. Moreover, this category turns out to 
include the bounded derived category $D^b(X)$ as a full subcategory, 
therefore the result is consistent with previous work on the subject 
\refs{\MRD, \AL}. We stress that we will not attempt to settle 
the question 
whether these two categories are equivalent. If they were equivalent, 
this would mean that the string field approach brings nothing new.
In that case, it would still be interesting to have an explicit 
construction of the equivalence. This section follows ideas proposed 
in section 5 of \CLiii\ and the general construction of \CLii.
The relation with twisted complexes 
of Bondal and Kapranov, enhanced triangulated categories, as well 
as the derivation of (an extension of) $D^b(X)$ from string field 
theory have been already discussed there. Since their discussion 
is brief, we spell out some details below. 

In order to avoid any technical complications, 
we will consider only Chan-Paton bundles of finite rank, as in \AL.\
Hence all but finitely many $E_n$ will be zero, and we are dealing with 
the bounded derived category $D^b(X)$. Note though that there is no 
convincing reason for this restriction from what has been said so far. 
In fact, from the point of view of string field theory it appears 
to be more natural to work with infinite complexes
(see also \refs{\KHi, \EWiv}.) We will not 
pursue this further in the present paper. 

We start by a giving a more formal description of the set of solutions 
to the equations of motion \eqmotionA,\ \eqmotionB.\ Recall that 
at the end of section 2, we have introduced a DG category $\CE$ 
whose objects are holomorphic vector bundles on $X$. The morphisms 
are given by 
\eqn\morphismsB{
\hbox{Hom}_{\CE}(E,F) = \oplus_{q=0}^3\Omega^{0,q}(E^\ast\otimes F),}
which is a graded abelian group with differential 
\eqn\diffB{
d_\CE = {\bar\del}_{E^\ast\otimes F}.} 
The cohomology of the morphism complex $\hbox{Hom}^{\bullet}_{\CE}(E,F)$ 
describes the physical operators  of topological open string theory 
in the presence of two D-branes $E$ and $F$. 

Bondal and Kapranov \BK\ introduced a formal construction 
which associates 
to any DG category an enlarged DG category 
whose objects are twisted complexes. 
This enhanced category has been denoted by $\pre(\CE)$ in \BK.\
We show below that their twisted complexes are formally identical 
to the solutions to \eqmotionB,\ although there are some sign
differences. A twisted complex is a 
collection $\{E_n\}$ of objects of $\CE$ with morphisms
$q_{mn} \in \hom^{m-n+1}(E_m, E_n)$ satisfying the equation 
\eqn\twistedA{
d_\CE q_{mn} + \sum_p q_{pn}q_{mp}=0.}
Given \morphismsB,\ it follows that the morphisms $q_{mn}$ 
are forms in $\Omega^{m-n+1}(E_m^\ast\otimes E_n)$. 
By comparing with \sfC,\ it follows that the $q_{mn}$ are in one to one 
correspondence with 
the components of the ghost number one string field in the presence 
of a graded collection of D-branes $\{E_n\}$. 
Moreover, the equations \twistedA\ are formally identical to the 
equations of motion \eqmotionB.\ The main difference is that 
in \twistedA\ the $q_{mn}$ are multiplied as ordinary forms whereas 
in \eqmotionB\ the $q_{mn}$ are multiplied as elements of the 
superalgebra $\CA$. In order to distinguish between the two products we 
will denote ordinary multiplication of differential forms by $\wedge$, 
as usual. 

Now let us describe the morphisms of $\pre(\CE)$, that is to each pair 
of objects $C=\{E_n, q_{mn}\}$, $C'=\{E'_n, q'_{mn}\}$, we associate 
a graded abelian group $\hom_{\pre(\CE)}(C,C')$, with a differential 
$d_{\pre(\CE)}$. 
We have \BK\
\eqn\twistedB{\eqalign{
& \hom^k_{\pre(\CE)}(C,C')=\oplus_{q+n-m=k}\hom^q_\CE(E_m, E'_n),\cr}}
and for $f_{mn}\in \hom^q_\CE(E_m, E'_n)$
\eqn\twistedC{
d_{\pre(\CE)}f_{mn} = d_\CE f_{mn} +\sum_{p}q'_{np}\wedge f_{mn} + 
(-1)^{q(m-p+1)} f_{mn} \wedge q_{pm}.}
This defines a DG structure on $\pre(\CE)$. 

In our case, the morphisms can be defined similarly, but we have to 
take into account the fact that the relevant algebra structure is $\CA$. 
This means the equation \twistedC\ has to be replaced by 
\eqn\twistedCA{
d_{\CQ}f_{mn} = Df_{mn} + \sum_{p} {q^\prime}_{np}f_{mn} -
(-1)^{l+n-m} f_{mn} q_{pm},}
where $D$ is the superconnection defined in \superF.\ 
This is a specialization of the general construction of \CLii\ 
to the case at hand. 
Although this construction looks rather complicated, let us note that 
it has 
a natural physical interpretation \CLii. 
We noticed before that the twisted 
complexes are nothing else than solutions to string field theory which 
define new deformed topological string theories. The deformations 
are encoded by the maps $q_{mn}$. Each deformation of the topological 
field theory should reflect in a deformation of the BRST operator, 
as in \refs{\MRD,\CLii,\AL}. The formula \twistedC\ defines the deformed 
BRST operator corresponding to two generalized D-branes defined by the 
objects $C, C'$. It acts on the graded space 
$\hom^{\bullet}_{\CQ}(C,C')$ which represents the space of 
open string 
states stretching between $C$ and $C'$. The grading is induced by ghost 
number. 

In order to complete the picture, we need to take one last step, 
namely to keep only the physical open string states. 
This can be achieved 
by passing to the cohomology category associated to $\CQ$. This means 
that we keep the same objects, but we replace the morphisms 
$\hom^{\bullet}_{\CQ}(C,C')$ by the graded abelian group 
$H\left(\hom^{\bullet}_{\CQ}(C,C')\right)$. 
In other words, 
we take BRST cohomology on the open string states, 
keeping only inequivalent 
physical states in each ghost degree. 
We will denote the resulting category by $H(\CQ)$. 

Strictly speaking, keeping all 
ghost degrees might be superfluous. It should suffice to restrict to 
cohomology of degree zero $H^0\left(\hom_{\CQ}(C,C')\right)$, 
in which case we obtain the category $H(\CQ)$ which is analogous 
to $\hbox{Tr}(\CE)$ defined in \BK. 
For example a similar phenomenon takes place for the
class of topological open string theories considered in \refs{\MRD,\AL}. 
That is, there is no loss of information if one keeps only the 
cohomology 
of degree zero of the D-brane category ${\bf T}(X)$ defined in \AL. 
The higher cohomology is recovered by applying the shift functor.
So we conjecture that the category encoding all the physical
information is $H^0(\CQ)$. 
It is known \BK\ that $\hbox{Tr}(\CE)$ is a triangulated category. 
A natural conjecture would be that the physical category $H^0(\CQ)$ 
is also 
triangulated in order to successfully describe decay phenomena as 
in \MRD. We will not attempt to prove this here.  

To conclude this section, let us investigate in some detail the relation 
between the enlarged D-brane category found in this section and 
the derived category $D^b(X)$. We will only be able to show that 
the derived category is equivalent to a full subcategory of $H^0(\CQ)$,
using the model of \AL\ for $D^b(X)$.  
To this end, let us consider the full subcategory of $\CQ$ 
generated by objects $C=\{E_n, q_{mn}\}$ which are complexes i.e. 
$q_{mn}=0$, unless $n=m+1$. In this case, the relations \twistedA\
reduce to 
\eqn\subcatB{\eqalign{
& {\bar \del} q_{m,m+1}= 0\cr
& q_{m,m+1} q_{m-1,m} = 0,\cr}}
which show that $C$ is a holomorphic complex of holomorphic vector 
bundles. 
Let us analyze the morphisms in $H(\CQ)$ between two such objects. 
For this, we have to specialize the formulae \twistedB,\ \twistedC\ 
to the case at hand, i.e. for two complexes $C, C'$
\eqn\specialization{\eqalign{
 \hom^k_{\CQ}(C, C') = &
\oplus_{q=0}^3 \oplus_{m} \Omega^{0,q}(E_m^\ast\otimes E'_{m+k-q})\cr
 d_{\CQ} f_{m, m+k-q} = &{\bar\del}f_{m, m+k-q}+ 
q'_{m+k-q,m+k-q+1}f_{m, m+k-q}-\cr
& (-1)^{k}f_{m, m+k-q}q_{m-1,m},\cr}}
where $f_{m, m+k-q}\in \Omega^{0,q}(E_m^\ast \otimes E'_{m+k-q})$.
This yields a differential complex, whose cohomology defines 
$\hom_{H(\CQ)}(C, C')$; the cohomology of 
degree zero defines $\hom_{H^0(\CQ)}(C, C')$. Let ${\bf S}(X)$ denote 
the full subcategory of $H(\CQ)$ generated by complexes;
${\bf S}_0(X)$ will denote the corresponding full subcategory of 
$H^0(\CQ)$. Since it will be needed shortly, it is convenient to rewrite 
the differential \specialization\ in terms of ordinary form 
multiplication
\eqn\ordmult{\eqalign{
& d_{\CQ}f_{m, m+k-q}= {\bar \del}f_{m, m+k-q}+\cr
&(-1)^k\left[(-1)^{q-k} q'_{m+k-q,m+k-q+1}\wedge f_{m, m+k-q}- 
f_{m, m+k-q}\wedge q_{m-1,m}\right].\cr}}

Our goal is to compare ${\bf S}(X)$, ${\bf S}_0(X)$ with  
${\bf T}(X)$, ${\bf T}_0(X)$ defined in \AL\ section 2. 
Recall that the objects of ${\bf T}(X)$ are holomorphic complexes of 
holomorphic vector bundles on $X$, therefore it has the same objects as 
${\bf S}(X)$. The morphisms of ${\bf T}(X)$ are graded abelian groups 
defined as the cohomology of the double complex
\eqn\doublecomplex{
\matrix{ & & {\bar \del} \bigg\uparrow & & 
{\bar \del} \bigg\uparrow & \cr
& {\buildrel {\bar q}\over \ra} & 
\Omega^{0,1}(\chom^0(E_\bullet, E'_\bullet))
& {\buildrel {\bar q}\over \ra} & 
\Omega^{0,1}(\chom^1(E_\bullet, E'_\bullet))
& {\buildrel {\bar q}\over \ra} \cr
& & {\bar \del} \bigg\uparrow & & {\bar \del} \bigg\uparrow & \cr
& {\buildrel {\bar q}\over \ra} & 
\Omega^{0,0}(\chom^0(E_\bullet, E'_\bullet))
& {\buildrel {\bar q}\over \ra} & 
\Omega^{0,0}(\chom^1(E_\bullet, E'_\bullet))
& {\buildrel {\bar q}\over \ra} \cr
 & & {\bar \del} \bigg\uparrow & & {\bar \del} \bigg\uparrow & \cr}}
where 
\eqn\homdef{
\chom^k(E_\bullet, E'_\bullet) = \oplus_{m} \chom(E_m, E'_{m+k}).}
The horizontal differential ${\bar q}$ can be taken to be 
\eqn\hrdiff{
{\bar q} f_{mn} = (-1)^{n-m}q'_{n, n+1}\wedge f_{mn} - f_{mn}\wedge
 q_{m-1, m},}
which is in fact identical to the nonderivative part of the second 
equation in \specialization.\foot{In fact the horizontal differential of 
\AL\ was written as 
${\bar q} f_{mn} =q'_{n, n+1} f_{mn} + f_{mn} q_{m-1, m}$, using 
conventions in which $q^\prime, q$ anticommute. If we treat 
$q^\prime, q$ as ordinary differential forms, there is an extra sign
as in \hrdiff.}
It is now a straightforward exercise 
to check that 
the complex defined in \specialization\ is the simple complex 
associated to the 
double complex \doublecomplex.\ This shows that they have isomorphic 
cohomology, therefore the categories ${\bf S}(X)$ and ${\bf T}(X)$ are 
equivalent. The same is true for ${\bf S}_0(X)$ and ${\bf T}_0(X)$. 
On the other hand, one of the main results of \AL\ is that 
${\bf T}_0(X)$ is equivalent to the derived category $D^b(X)$. 
Hence we have effectively identified $D^b(X)$ with a full subcategory 
of $H^0(\CQ)$. 

{\centerline{\bf Acknowledgments}}
I am very grateful to Michael Douglas for many explanations on his 
work and for sharing his insights and ideas with me, and Jaume Gomis, 
Greg Moore, and especially Paul Horja for very useful discussions 
and encouragement. I thank Paul Aspinwall, Michael Douglas, 
Calin Lazaroiu and Eric Sharpe
for comments on the first version of the paper. 
I would also like to acknowledge the 
support of DOE grant 
DE-FG02-90ER40542, and the hospitality of the ITP Santa Barbara
where part of this work was done.

\listrefs
\end